\documentclass[referee]{raa}            

\usepackage{graphicx,times}             
\usepackage{natbib}
\usepackage{amssymb,amsmath}
\usepackage{float}
\bibpunct{(}{)}{;}{a}{}{,}

\usepackage[a4paper=true,dvipdfm=true,pagebackref=true]{hyperref}
\hypersetup{colorlinks = true, linkcolor = green, anchorcolor = red, citecolor = blue, filecolor = red, pagecolor = red, urlcolor = red}

\begin{document}

   \title{Luminosity of radio pulsar and its new emission death line
}

   \volnopage{Vol.0 (20XX) No.0, 000--000}      
   \setcounter{page}{1}          

   \author{Q. D. Wu
      \inst{1,3}
   \and Q. J. Zhi
      \inst{1,3}
   \&  C. M. Zhang
      \inst{2,4}
   \and D. H. Wang
      \inst{1,3}
   \and C. Q. Ye
      \inst{1,3}
   }

   \institute{School of Physics and Electronic Sciences, Guizhou Normal University, Guiyang 550001, China; {\it qjzhi@gznu.edu.cn, 17010070218@gznu.edu.cn}\\
        \and
             National Astronomical Observatories,Chinese Academy of Sciences, Beijing 100012, China; zhangcm@bao.ac.cn\\
        \and
             Guizhou Provincial Key Laboratory of Radio Astronomy and Data Processing, Guiyang 550001, China\\
        \and
             University of Chinese Academy of Sciences, Beijing 100049, China\\
\vs\no
   {\small Received~~06-Mar-2020; accepted~~30-May-2020}}

\abstract{ We investigated the pulsar radio luminosity (L), emission  efficiency (ratio of radio luminosity to its spin-down power $\dot{E}$), and death line in the
diagram of magnetic field ($B$) versus spin period ($P$), and
found that the dependence of pulsar radio luminosity on its spin-down power ($L - \dot{E}$) is  very weak,
shown as $L \sim \dot{E}^{0.06}$, which deduces an
equivalent  inverse correlation between  emission  efficiency and    spin-down power  as  $\xi \sim \dot{E}^{-0.94}$.
%
%
Furthermore, we examined the distributions of radio luminosity of millisecond and normal pulsars, and found that, for the similar spin-down powers, the radio luminosity of   millisecond pulsars is about one order of magnitude lower than that of the normal pulsars.
The analysis of pulsar radio flux suggests that this correlations are not due to a selective effect, but are intrinsic to the pulsar radio emission physics. Their radio radiations  may be dominated by the different radiation mechanisms.
The  cut-off phenomenon of currently observed  radio pulsars in   $B-P$ diagram is usually referred as the ``pulsar death line'',
which corresponds to $\dot{E}\approx10^{30}$\,erg$\cdot$s$^{-1}$ and is obtained by  the cut-off voltage of electron acceleration gap
in the polar cap model of pulsar proposed  by Ruderman and Sutherland.
Observationally, this death line can be inferred by
the actual observed pulsar flux $S\geq$\,1\,mJy and 1\,kpc distance, together with the maximum radio emission efficiency of
1\%.
However, the observation data show that the 37 pulsars pass over the death line, including the recently
observed two pulsars with long periods of  23.5\,s and 12.1\,s ,
which violate the prediction of polar cap model.
At present, the actual observed pulsar flux can reach 0.01\,mJy by FAST telescope,  which will arise  the observational
limit of spin-down power of pulsar as low as  $\dot{E}\approx10^{28}$\,erg/s. 
This means that the new death line is downward shifted two orders of magnitude, which might be favorably  referred as the ``observational limit--line'',
and accordingly the pulsar theoretical  model for the cut-off voltage of gap should be heavily modified.
\keywords{stars: neutron --- (stars:) pulsars: general ---stars: fundamental parameters}
}

   \authorrunning{Q. D. Wu, Q. J. Zhi \& C. M. Zhang }            

   \maketitle

%
%
\section{Introduction}           
\label{sect:intro}

\begin{figure}[h]
\center{\includegraphics [width=10cm]{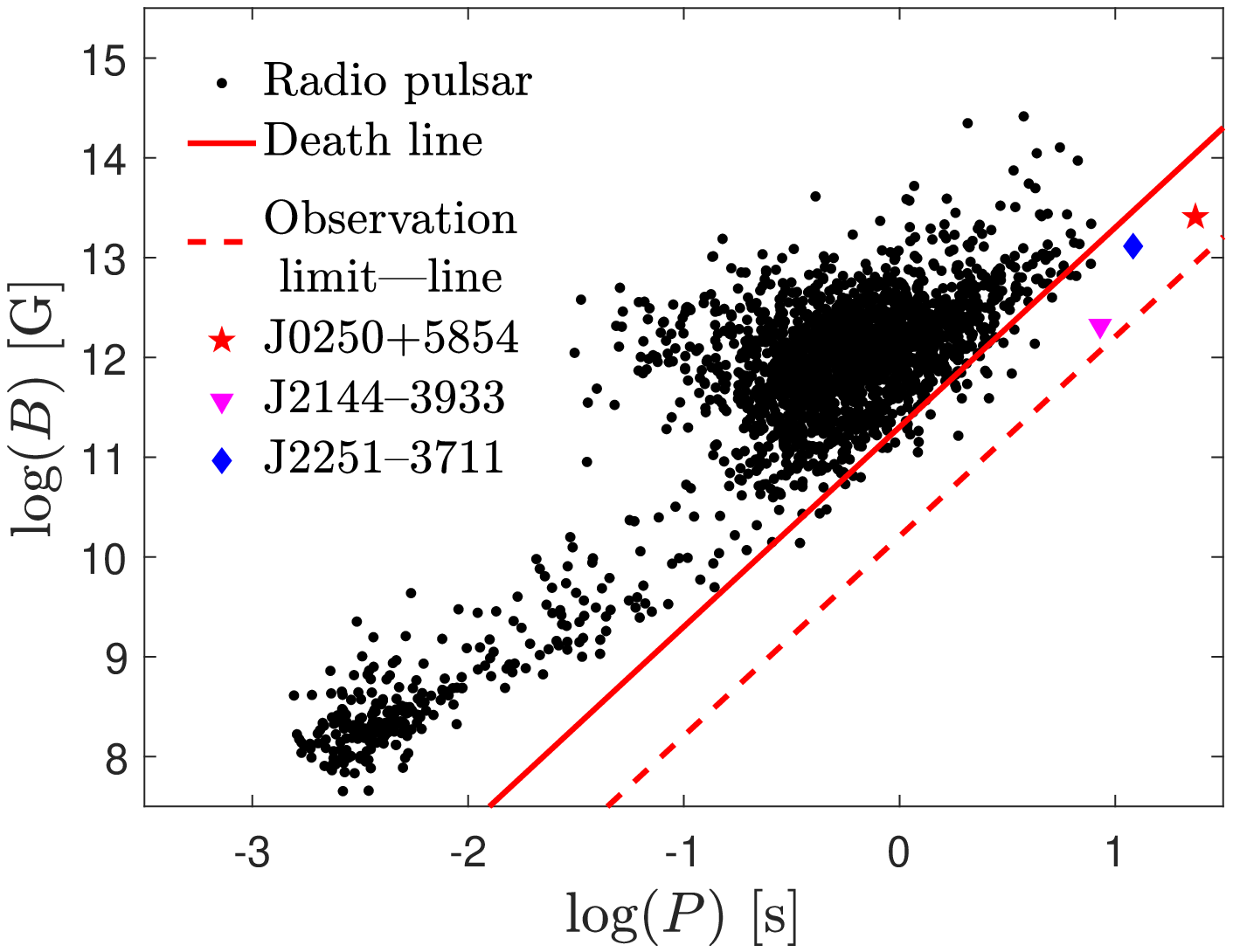}}
\caption{The magnetic field versus period ($B-P$) diagram for the radio pulsars. The solid line is the death line defined as $B_{12}/P^{2}\cong0.2$ \citep[e.g.,][]{1975ApJ...196...51R,1991PhR...203....1B}, and the dashed line is the ``observational limit---line'' corresponds to $\dot{E}\simeq10^{28}$\,erg\,s$^{-1}$ (see section~\ref{sec:2.4}). The pulsar samples are taken from the ATNF Pulsar Catalogue:   https://www.atnf.csiro.au/research/pulsar/psrcat/ \citep{2005AJ....129.1993M} }
\label{fig1}
\end{figure}

It is generally accepted that the radio radiation of a pulsar originates from the generation of electron-positron pairs in its magnetic magnetosphere \citep{1975ApJ...196...51R,1971ApJ...164..529S,1969ApJ...157..869G,1979ApJ...231..854A,van den Heuvel(2006),1979STIN...8025219M,1980ApJ...235..576C,1978ApJ...225..557M,1997ARepS...1..187L}.
When the radio pulsar does not have enough pairs to generate the pulses, it will be ceased \citep[][hereafter RS75]{1975ApJ...196...51R}, which is ascribed to the limited voltage of the gap $\Phi_{max}\approx\frac{BR^{3}\Omega^2}{2c^2}$, where $B$ is the neutron star (NS) polar  magnetic field strength at surface, $\Omega$ is the angular velocity associated with spin period $P$ as $\Omega=2\pi/P$, and $R$ is the NS radius.
The so-called radio pulsar death line is defined by setting $\Delta V=\Phi_{max} = 10^{12} {\rm Volt } \Rightarrow B_{12}/P^2\simeq0.2$ \citep{1991PhR...203....1B}, where $B_{12}$ is the NS magnetic field in the units of $10^{12}$\,G \citep{1975ApJ...196...51R,1991PhR...203....1B}.
Under the assumption of the magnetic dipole model \citep{1983bhwd.book.....S,2012hpa..book.....L,2012puas.book.....L}, the limited voltage of the gap corresponds to the rotational energy loss rate of $\dot{E}\simeq1.5\times10^{30}$erg$\cdot $s$^{-1}$ \citep{1991PhR...203....1B}, which is usually referred as the death line of radio pulsars, as shown in Fig.\ref{fig1}.
Until now approximately 2700 radio pulsars have been found \citep[ATNF Pulsar Catalogue,][]{2005AJ....129.1993M,2019BAAS...51c.261L}, most of which are situated above the death line, in the magnetic field versus period ($B-P$) diagram (see Fig. \ref{fig1}).
However, from Figure.~\ref{fig1}, we find that 37 pulsars are observed below the death line, including the long-period pulsar J0250+5854  \citep[$P=23.5$\,s,][]{2018ApJ...866...54T} and J2251--3711\citep[$P=12.1$\,s,][]{2019arXiv191004124M} and J2144--3933 \citep[$P=8.5$\,s,][]{1999Natur.400..848Y}.
These pulsars are characterized as low spin-down  power radio pulsars (hereafter low-$\dot{E}$), thus
the RS75 model should be modified to explain this fact.

\citet{2000ApJ...531L.135Z}  reinvestigated  the radio pulsar ``death lines'' within the framework of the vacuum gap model (V---model) and the space---charge---limited flow model (SCLF) with either curvature radiation (CR) or inverse Compton scattering (ICS) photons as the source of pairs. They found that the ICS induced SCLF model can maintain a strong pair generation in the pulsar J2144--3933.
\citet{2018ApJ...866...54T} found that the curvature radiation and inverse Compton scattering death lines of  SCLF model \citep{2000ApJ...531L.135Z} were both located below the position of PSR J0250 + 5854.
\citet{2001ApJ...550..383G} argued that the death line of curvature radiation can be moved further down by considering very curved magnetic field lines with a radius of curvature much smaller than the radius of a typical neutron star.
\citet{2011ApJ...726L..10H} proposed an offset pole with a distorted magnetic field. The death line of curvature radiation in the SCLF model can move downward.
\citet{2017MNRAS.472.2403Z} investigated the neutron star equation of state and found that the heavier neutron stars can explain the presence of radio pulsars outside the standard death line.
The most concise explanation of the current death line is proposed by \citet{2014ApJ...784...59S}.
They believe that the upper limit of radio radiation efficiency ($\xi\sim0.01 $) is equivalent to the death line.
Although there are many explanations at present, none of them can explain the fact that 37 radio pulsars pass through the death line.

In this paper, to pursue the cause of the death line crisis, we investigated the  luminosity, radiation efficiency, and ``death line'' of radio pulsars.
In Section~\ref{sec:2}, we introduce the calculation formula of radio luminosity and the definition of radiation efficiency.
And explore whether the $\xi-\dot{E}$ association is intrinsic, where  the effects of pulsar magnetic field and
period on radiation efficiency are studied, and we present the ``observational limit---line'' to replace the death in $B-P$ diagram.
Sections~\ref{sec:3}  is dedicated to the discussion and conclusion.

\section{statistics of pulsar radio luminosity}\label{sec:2}

In this section, we study the relation between the spin-down power and radio luminosity of pulsars.

\subsection{Radio luminosity and emission efficiency}\label{sec:2.1}

The  precise estimation of the pulsar radio luminosity   is difficult  on account of  various reasons \citep[see][]{2012hpa..book.....L},
and our analysis is based on the  radio pulsar  luminosities of ATNF Catalogue  at the wave band of  1400\,MHz \citep{2005AJ....129.1993M}.
The observed flux density of a radio source is measured in Jansky defined as  1\,Jy=$10^{-26} W\cdot m^{-2}\cdot$ Hz$^{-1}=10^{-23}$erg$\cdot $s$^{-1}\cdot$ cm$^{-2}\cdot $Hz$^{-1}$, based on which    the total radio luminosity ($L$) of the pulsar is calculated by the formula provided by   \citet{2012hpa..book.....L}:
\begin{equation}
   L = \frac{4\pi d^2}{\delta} sin^2(\frac{\rho}{2})\int_{\nu_{min}}^{\nu_{max}} S_{mean}(\nu)\,d(\nu)
    \label{eq1}
    \end{equation}
where $d$ is the pulsar distances, the pulse duty cycle $\delta=W_{eq}/P$ with $W_{eq}$ being  the equivalent pulse width.
$\nu_{min}$ and $\nu_{max}$ describe the frequency range in which the pulsar is detected and studied, $S_{mean}(\nu)$ is the mean flux density measured at frequency $\nu$, and $\rho$ is the   emitting angle of the pulse beam.
\citet{2012hpa..book.....L} employ 1400\,MHz as the reference frequency and assume  the typical values for all pulsars: $\delta\approx0.04$, $\rho\approx6^{\circ}$, $\nu_{min}\approx10^7$\,Hz, $\nu_{max}\approx10^{11}$\,Hz, thus the equation~\ref{eq1} can be expressed as follows:
\begin{equation}
    L\simeq7.4\times10^{27}\,{\rm erg\cdot s^{-1}}(\frac{d}{\rm kpc})^{2}(\frac{S_{\rm1400}}{\rm mJy})
    \label{eq2}
    \end{equation}
where $S_{1400}$ is the mean flux density at 1400\,MHz\,(mJy).

For the efficiency of radio pulsar  radiation, defined as ratio of  radio emission power to that of NS  spin-down power,
can be written as  \citep[e.g.,][]{2014ApJ...784...59S,2006ARep...50..483M},

\begin{equation}
    \xi\equiv\frac{L}{\dot{E}}
    \label{eq3}
    \end{equation}
where $\dot{E}$ is the spin-down power (also called spin-down luminosity).

\subsection{Radio luminosity and spin-down power}\label{sec:2.2}

\begin{figure}[h]
\center{\includegraphics [width=14cm]{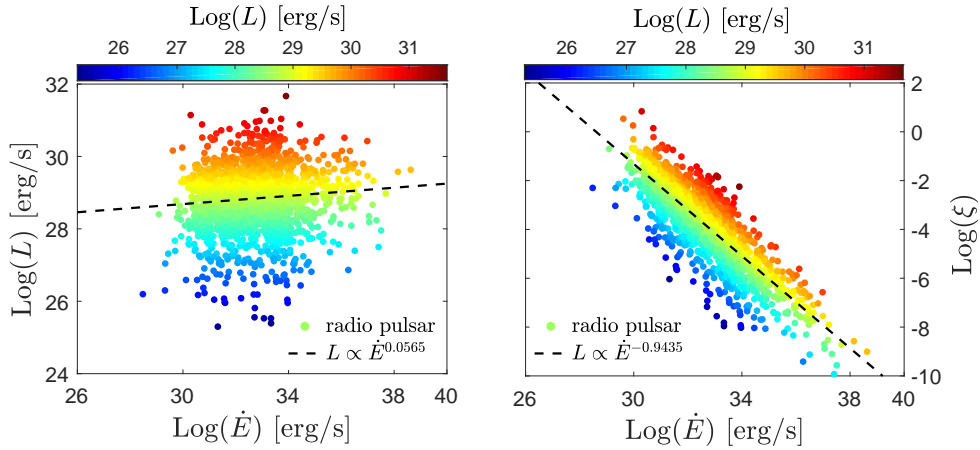}
        }
\caption{Dependence of the radio luminosity (left panel) and  emission efficiency (right panel) on the spin-down power. The colors of samples correspond to the different values of radio luminosity. The pulsar samples are taken from the ATNF Pulsar Catalogue: https://www.atnf.csiro.au/research/pulsar/psrcat/ \citep{2005AJ....129.1993M}.  }
\label{fig2}
\end{figure}

For radio pulsars, to uncover their emission mechanisms, we investigate the statistical properties of the radio luminosity ($L$) and emission efficiency ($\xi$) as a function of the spin-down power, respectively, as shown in Figure. \ref{fig2}.
It is worth noting that the quantities $L$ and $\dot{E}$ of the pulsar are  two independent parameters by the  different measurements.
The best fitting indicates  that there is  basically a very weak  dependence between the  radio luminosity and spin-down power, as $L\propto\dot{E}^{0.0565}$.
The fitting coefficients of the power law index (with 95\% confidence bounds) is given as
 0.0565 with the regime of  (0.0296,0.0834), implying that the correlation is still weak.
In other words,  the correlation is very diffuse and the goodness (R-square) of fit is only 0.093.
In addition, the dependence of radio emission efficiency and spin-down power,  $\xi-\dot{E}$, as shown in Figure. \ref{fig2},
is also  found  to be $\xi\propto\dot{E}^{-0.94}$.
 In fact, both correlations $L - \dot{E}$ and  $\xi-\dot{E}$ are equivalent, since the difference of  their power-law indices
  is 0.06 - 0.94 = 1.  Therefore, the declaim of the inverse relation between the radio efficiency and spin-down power present
  no useful information of the pulsar intrinsic emission physics \citep{2014ApJ...784...59S}, or mathematically  the $\xi-\dot{E}$ correlation
   is equivalent to the $L-\dot{E}$ correlation.
An alternative interpretation of the process underlying the cessation of pulsar emission is presented in \citet{2014ApJ...784...59S}, where a model-independent statement can be made that the death line of radio pulsars corresponds to an upper limit in the efficiency of radio emission.
By the weak correlation  between $L - \dot{E}$ and  $\xi-\dot{E}$,  we know that this interpretation of pulsar radio signal ``death'' is unreliable.

\subsection{Radio luminosity  of recycled and normal pulsars}\label{sec:2.5}
 \begin{figure}[h]
\center{\includegraphics [width=14cm]{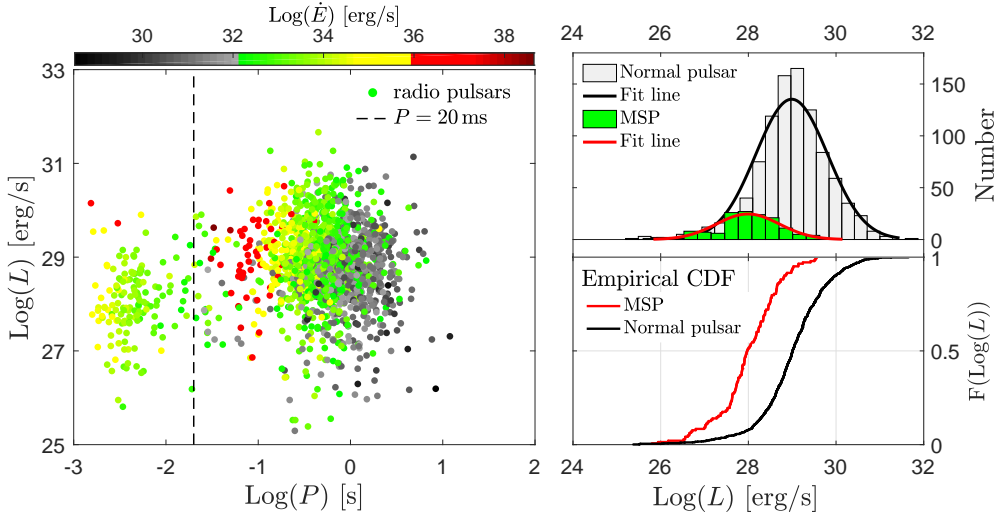}
       }
\caption{The left  panel shows the dependence of radio luminosity on the spin period. The left panel is the histogram of radio luminosity (upper) and cumulative distribution (lower). various  colors of samples correspond to the different values of pulsar spin-down power. }
\label{fig4}
\end{figure}
As can be seen from the left panel of Figure.~\ref{fig4}, the recycled pulsars and normal pulsars  can be divided into two parts by the boundary value of  $P=20$ ms, and the  radio luminosity of two types of pulsars have no clear correlations with the spin period.
However, we also noticed that the radio luminosity of millisecond pulsars is averagely lower than that of normal pulsars.
 For the  similar  spin-down power regimes of  $1.41\times10^{32}\sim6.25\times10^{35}$\,erg/s, the average values of both types are $3.02\times 10^{28}$\,erg/s and $4.84\times 10^{29}$\,erg/s, respectively, or the averaged  radio luminosity of millisecond pulsars is about one order of magnitude lower than that of the  normal pulsars.
A further comparison of normal and millisecond pulsars is performed  within the spin-down power range of $1.41\times10^{32}\sim6.25\times10^{35}$\,erg/s.
A cumulative distribution test of their radio luminosities also exhibited the same conclusion (see CDF in Figure.~\ref{fig4}).
Through the histogram test, it was found that the radio luminosity distribution of both types of  pulsars have two distinct peaks (see the histogram in Figure.~\ref{fig4}).
These clearly express the significant difference between the two types of pulsars, which may provide the valuable information for the distinctive pulsar's radio radiation mechanism of both types of  pulsars.

\subsubsection{Selective effect}\label{sec:2.5.1}

 \citet{1997ApJ...482..971C} likelihood analysis on the data from extant surveys (22 pulsars with spin periods less than 20 ms) accounts for the following important selection effects: (1) the survey sensitivity described by the flux density ($S_{\rm min}$) as a function of direction, spin period, and sky coverage; (2) the interstellar scintillation, which modulates the pulse flux and causes a net increase in search volume of ~30\%; and (3) errors in the pulsar distance scale.
It can be seen that its selection effect is mainly affected by the survey sensitivity and pulsar flux density.
The detectability of pulsars by the work of  \citet{2006ApJ...643..332F}   depends on its inherent characteristics (such as brightness, pulse period and duty cycle), its location (distance, DM, interstellar scattering and brightness temperature of the background sky) and the details of the observation system.
These factors all affect the minimum observable flux density ($S_{\rm min}$).
We  studied  the pulse flux density distribution of millisecond pulsars and normal pulsars, and then analyzed  whether there is a selective effect.
\begin{figure}[h]
\center{\includegraphics [width=10cm]{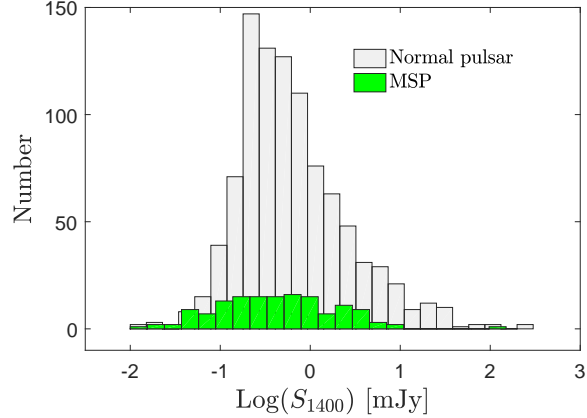}}
\caption{Histogram of flux density at 1400\,MHz.}
\label{figs}
    \vspace{-0.3cm}
\end{figure}
Under the condition that the energy loss rate range is $1.41\times10^{32}\sim6.25\times10^{35}$\,erg/s,  the observed flux density is shown in Figure.~\ref{figs}, where the distribution range and peak value of flux density for both types of pulsars  are consistent, so the selection effect is not found.
Therefore, the analysis of radio flux density suggests that the upper part correlations are
not due to the selection effect, but are intrinsic to the   radio emission physics of pulsars.

\subsubsection{emission efficiency}\label{sec:2.5.2}
\begin{figure}[h]
\center{\includegraphics [width=14cm]{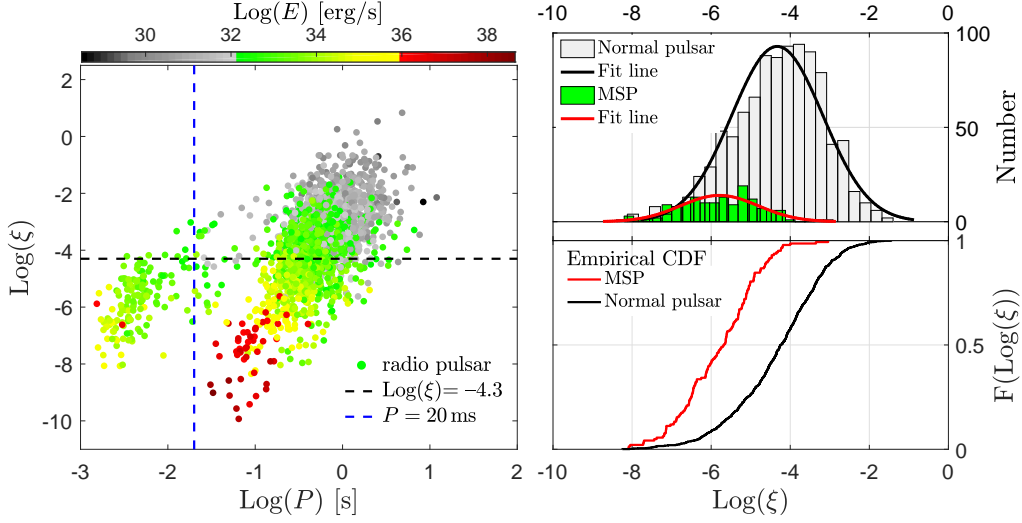}}
\caption{The left  panel shows the dependence of radio emission efficiency on the spin  period. The right  panel is the histogram of radio emission efficiency (upper) and cumulative distribution (lower). The colors of samples correspond to the different values of pulsar spin-down power.}
\label{figp-bi}
    \vspace{-0.3cm}
\end{figure}
As can be seen from the upper panel of Figure.~\ref{figp-bi},  the radio emission efficiency of the pulsar is positively
correlated with the spin period (for $P>20 $ms with $L\propto P^{1.9}$), which was noticed by \citet{2006ARep...50..483M}. However, this conclusion
has little physical significance, since the correlation $\dot{E}\propto\dot{P}/P^3$ will result in the proportional dependence of the efficiency $\xi$ to the spin period.
We noticed that the radio emission efficiency of millisecond
pulsars is generally less than $10^{-4}$ (see in Figure.~\ref{figp-bi}), which is lower than those   of most  normal pulsars.
For a similar spin-down power range $1.41\times10^{32}\sim6.25\times10^{35}$\,erg/s.
Their average radiation efficiency is $1.73\times10^{-5}$ and $5.69\times10^{-4}$, respectively.
Thus, it is found that the emission efficiency of millisecond pulsars is averagely one  order of magnitude lower than that of normal pulsars.
Examination of these two types of pulsars by histogram and cumulative distribution reveals that there exists  a relatively large difference in their emission efficiency.
This provides us with very valuable information because,  for some physical reason, the radio luminosity of the millisecond pulsar at a similar spin-down power is one order of magnitude lower than that of a normal pulsar, and ultimately its emission efficiency is also one order of magnitude lower, which  provides a valuable hint  for the different  radio radiation process of two types of pulsars.

\subsection{radio fluxes and spin-down power relation}\label{sec:2.3}

\begin{figure}[h]
\center{\includegraphics [width=14cm]{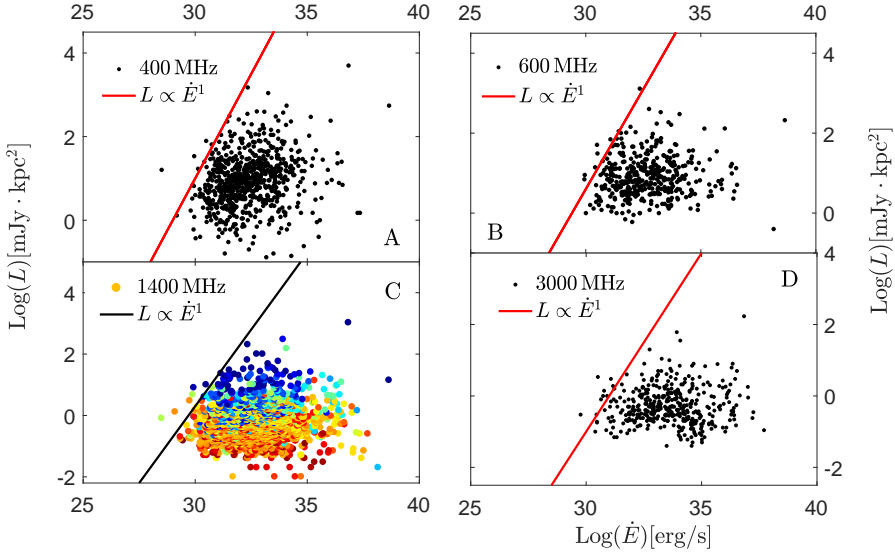}}
\caption{Dependence of the flux density on the spin-down power for various radio bands. The sub-figures A, B, C and D show the $S - \dot{E}$ relationship at center frequencies 400\,MHz, 600\,MHz, 1400\,MHz, and 3000\,MHz, respectively. The color of the sample in the  sub-figure  C  corresponds to the date of pulsar discovery.}
\label{fig3}
    \vspace{-0.3cm}
\end{figure}
We investigate the statistical properties of the radio mean flux density as a function of the spin-down power as shown in Figure.~\ref{fig3}.
 There is basically a very weak dependence between the two quantities.
When the spin-down power of the pulsar is reduced to $\sim\dot{E} \simeq 10^{34}$ erg/s. We notice that the maximum flux density of pulsars increase with  the spin-down power. The impact of energy loss rate on the pulsar flux density exist  significantly.
We found that there exists   a proportional relation $S\propto\dot{E^1}$  between the flux density and spin-down power, which is clear
  when the spin-down power is less than $\dot{E} \simeq 10^{34}$ erg/s.
This provides us with  very interesting and valuable information on  radio radiation mechanism.
Furthermore, the relation between the flux density and spin-down power has no significant cut-off near $\dot{E} \simeq 10^{30}$ erg/s.
In addition,
it is worth noting  that the value of the flux density has a significant stratification in the date of pulsar
discovery, or more recently discovered pulsars are shown with the low flux density (see in Figure.~\ref{fig3} at frequencies 1400\,MHz  ).

\subsection{Pulsar distance}\label{sec:2.6}

\begin{figure}[h]
\center{\includegraphics [width=8.6cm]{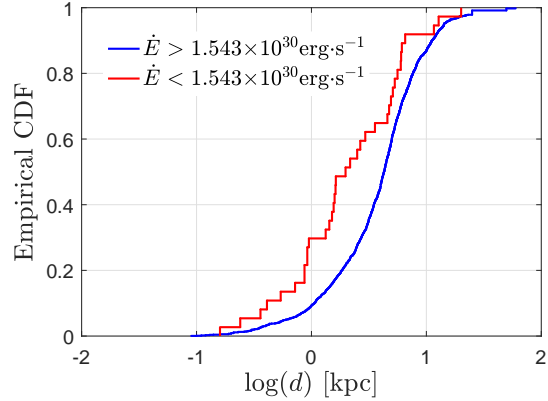}}
\caption{The cumulative distribution of pulsar distance. The red line belongs to the low-$\dot{E}$ radio pulsar, and the purple line is the other radio pulsar. }
\label{fig6}
\end{figure}

There are 37 low-$\dot{E}$ pulsars,  which are distributed below the death-line ($\dot{E} \simeq 10^{30}$ erg/s) in $B-P$ diagram (see  Figure.~\ref{fig1}). To investigate their
properties, we compare their distances to our solar system with the other pulsars that are distributed above the death-line.
In the Figure.~\ref{fig6}, we find that  the distances of the low-$\dot{E}$ pulsars are  relatively close  to  our solar system,
 compared to those of other pulsars.
For two groups of pulsars, below/above the death-line that are corresponding to low/high $\dot{E}$,  we study  their distance distributions by
Kolmogorov-Smirnov test (hereafter KS-test),  and obtain that  the returned value of h = 1
 rejects the null hypothesis at the default 5\% significance level, implying the two groups belong to the different distributions.
  In detail,  the average distance of  the low (high) $\dot{E}$ pulsars  is 3.6\,kpc  (5.6\,kpc).

\subsection{Death line and  ``observation limit-line'' } \label{sec:2.4}

\begin{figure}[h]
\center{\includegraphics [width=14cm]{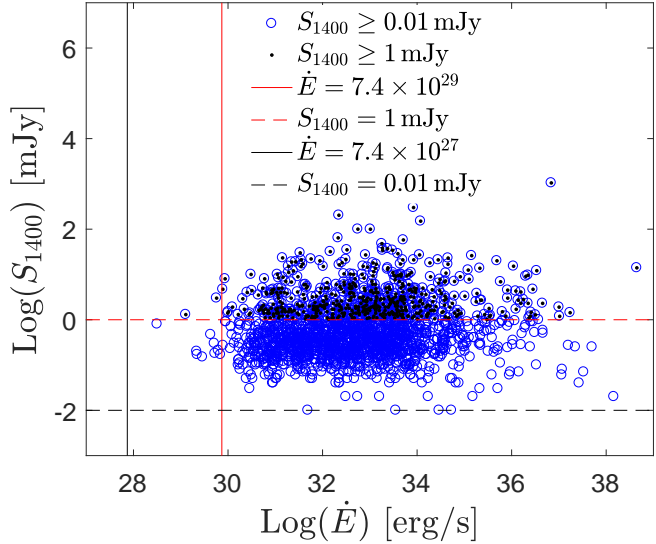}}
\caption{Dependence of the flux density on the spin-down power. }
\label{figs-e}
\end{figure}

Considering equation~\ref{eq2} and  equation~\ref{eq3}, we can get a formula of the radio efficiency,
\begin{equation}
    \xi\equiv\frac{L}{\dot{E}}\simeq\frac{7.4\times10^{27}\,{\rm erg\cdot s^{-1}}(\frac{d}{\rm kpc})^{2}(\frac{S_{\rm1400}}{\rm mJy})}{\dot{E}}\;.
    \label{eq4}
    \end{equation}
Under the condition that the pulsar distance is 1\,kpc and   the  observed pulsar flux density  can  reach $\geq$ 1\,mJy,
 the corresponding  radio luminosity
 of  $L\geq7.4\times10^{27}$\,erg$\cdot$s$^{-1}$ is obtained \citep{2012hpa..book.....L}. If the upper limit of
  pulsar radiation efficiency (equation~\ref{eq3}) is set as $\xi=$ 0.01  \citep[proposed by][]{2014ApJ...784...59S}, then  the corresponding spin-down power $\dot{E} \geq7.4\times10^{29} \approx10^{30}$\,erg$\cdot$s$^{-1}$ can be derived (see in Figure~\ref{figs-e}).
 This means that the minimum spin-down power that can be observed under this condition is $\approx10^{30}$\,erg/s.
This value is very close to the spin-down power of the   death line as $1.542\times10^{30}$\,erg/s \citep{1975ApJ...196...51R}.
At present, especially in the  observation of FAST (Nan et al. 2011), the actual observed pulsar flux is  about  $\geq0.01$\,mJy \citep{2005AJ....129.1993M}, which  is  corresponding to a radio luminosity of  $L\geq10^{26}$\,erg$\cdot$s$^{-1}$ at the distance of 1 kpc.
Therefore, the current ``observation limit---line'' can  reach  the order of $\dot{E}=10^{28}$\,erg$\cdot$s$^{-1}$ (see in Figure.~\ref{fig1} and Figure.~\ref{figs-e}).
 As shown in  Figure.~\ref{fig1},
it is obvious that all currently observed  radio pulsars are above this limit line, and as a result
the conception of pulsar graveyard \citep{1991PhR...203....1B}  is only that is beyond ``observation limit---line''  of the radio telescope.
This observation limitation   will lead to the phenomenon that the cutoff line of  ``seeing"" pulsars  happens  at a certain spin-down power value (see    Figure.~\ref{fig1}).

From the model construction, the researchers predict the death line of pulsar that is corresponding to the cut-off of the
polar cap for radio pulse emissions, however 37 low-$\dot{E}$ pulsars break this condition.
On the existence of the death line, we suspect its reality, on which the following arguments are proposed.

At first, from the $L - \dot{E}$ dependence, the pulsar luminosity is  very weak dependent of  $\dot{E}$, this means that the low-$\dot{E}$ does not means
a low radio luminosity.

Secondly, the 37 low-$\dot{E}$ pulsars are located very close to the solar system, which should account for their radio flux density  over the
telescope sensitivity.

Thirdly, from the flux density versus $\dot{E}$ diagram, there does not exist a sharp cut-off line to distinguish the low-$\dot{E}$ from the other sources,
but the low-$\dot{E}$ pulsars are randomly distributed.  From Figure.~\ref{fig3}, the maximum flux density of pulsars increase  with the   $\dot{E}$, so the low-$\dot{E}$
pulsars require the high sensitivity.

Fourth, the minimum observable spin-down power is related to the minimum value of the actual flux density currently observed.

Therefore, we think that it is the telescope sensitivity not the $\dot{E}$ to account for the less pulsars appearing below the ``death line",
 so enhancing
the telescope sensitivity could increase the number of pulsars there.
How the radio radiation of pulsars died is not clear, but we found that the classic death line (RS75) and the current pulsar's cut-off line can be ascribed  to the "observation limit--line".
Thus, we rather insist that there exists
a temporary ``observation limit--line", which  would be shifted to the low-$\dot{E}$ with enhancing the radio telescope sensitivity.

\section{Discussion and conclusion}\label{sec:3}

We investigate the statistical properties of radio luminosity of large samples of pulsars,  the following conclusions are drawn below.

(1) An alternative interpretation of the process underlying the cessation of pulsar emission is not reliable.
 Because mathematically, the  efficiency and spin-down power $\xi-\dot{E}$ correlation is equivalent to the $L-\dot{E}$ correlation, thus
 the inverse relation of $\xi-\dot{E}$ has the same meaning of the weak dependence of  $L-\dot{E}$.

(2) For the same range of $\dot{E}$ ($1.41\times10^{32}\sim6.25\times10^{35}$\,erg/s), on the radio luminosity of milliseconds and normal pulsars, we find that the average value of the latter is one order of magnitude higher than the average value of the former.
Their radio radiation is may dominated by different radiation mechanisms.
The analysis of radio fluxes suggests that this correlations are not due to a selection effect, but are intrinsic to the pulsar radio emission physics.

(3) The maximum flux density of pulsars increase with the $\dot{E}$, shown as  $S\propto\dot{E^1}$.. This relation  is evident when the spin-down power is less than $10^{34}$\,erg/s.

(4) From the RS75 model for the  radio pulsar emission, there exists a death-line in $B-P$ diagram as inferred from the cut-off voltage
of pulse emissions, which is corresponding to $\dot{E}\approx 10^{30}$\,erg/s.
From the pulsar observation, when the actual observed pulsar flux density  $S\geq$\,1\,mJy and distance of 1\,kpc  arises  a radio luminosity of $L \geq 10^{28}$\,erg/s, which  can deduce a limit $\dot{E}\approx 10^{30}$\,erg/s if the radio emission efficiency is 1\%.
At present, the actual observed pulsar flux density can reach a low value of 0.01\,mJy,  and the   current observation limit line is obtained as  $\dot{E}\approx10^{28}$\,erg$\cdot$s$^{-1}$  for the distance of 1 kpc and  maximum radio efficiency of 1\%, based on  which
 all observed pulsars in $B-P$ diagram lie above this ``new death-line".
 In other words, because of this limit line, with the enhance of  the telescope  sensitivity, pulsars with the spin-down power less than the
  limited value $\dot{E}\approx10^{28}$\,erg$\cdot$s$^{-1}$  can be observed (c.f. the cases of pulsars with the long spin periods of P=8.5s and  P=23.5 s).
Furthermore, the conception of the current death line  should be heavily modified, so does the proposed pulsar emission mechanism
  by RS model.

 Thus, the coincidence of theoretical minimum $\dot{E}\approx 10^{30}$\,erg/s  with the assumed minimum observational luminosity strengthens the existence of death line.  However,  the
37 pulsars are found to violate the condition of pulsar death, including the three longest period pulsar (J0250+5854, J2144--3933, and J2251--3711), investigations of which
indicates that most of the  37 pulsars are  detected with the lower sensitivity than 1 mJy, and with the averaged distance less than those above the death line.
The low spin-down powers are corresponding to the   long-period pulsars, however, the observations of which exist the selective effect.
As declaimed, due to the radio interference,  hardware and software high-pass filters, the  radio pulsar surveys typically reduce
the sensitivity to the long-period pulsars \citep{2006ApJ...643..332F}, and the presence of red noise also reduces the sensitivity
  detecting the  long-period pulsars \citep{2018ApJ...866...54T}.
  In addition,  most pulsar surveys last a relatively short  time of only a few minutes, which make most of
   long-period pulsars  missed if  there are few  pulses during the observations \citep{2018ApJ...866...54T}.
With the improvement of the detection efficiency for the   long-period pulsars, the more pulsar samples will appear in the
region of low-$\dot{E}$ in $B-P$ diagram.

\section{Acknowledgements}

This research is  Supported by National Natural Science Foundation of China (U1731238, 1731218, 11565010),The Science and Technology Fund of Guizhou Province ((2015)4015,(2016)-4008,(2017)5726-37), the National Natural Science Foundation of China NSFC (11773005, U1631236, U1938117. 11703001, 11690024, 11725313 ), NAOC-Y834081V01, the National Program on Key Research and Development Project (Grant No.2017YFA0402600), the Strategic Priority Research Program of the Chinese Academy of Sciences (Grant No.XDB23000000), and the CAS International Partnership Program (No.114A11KYSB20160008). {\bf  We thank R.N. Manchester, G. Hobbs and D. Lorimer for discussions. We are also grateful for the  anonymous referee for critic comments that help us to improve the quality of the paper. )

\label{lastpage}

\end{document}